\documentclass[useAMS]{mn2e}
\usepackage{psfig}
\usepackage{graphicx}
\usepackage{amsmath}
\usepackage{tabularx}

\usepackage{times}
\usepackage{amssymb,amsmath}

\def\lsim{ \lower .75ex\hbox{$\sim$} \llap{\raise .27ex \hbox{$<$}} }
\def\gsim{ \lower .75ex \hbox{$\sim$} \llap{\raise .27ex \hbox{$>$}} }

\title[The EC/CMB model revisited] 
{Revisiting the EC/CMB model for extragalactic large scale jets} 

\author[Lucchini, Tavecchio \& Ghisellini]
{M. Lucchini$^1$\thanks{E--mail: matteo.lucchini@brera.inaf.it}
F. Tavecchio$^1$ and G. Ghisellini$^1$\\
$^1$INAF -- Osservatorio Astronomico di Brera, via E. Bianchi 46, I--23807
Merate, Italy
}

\begin{document}



\maketitle

\begin{abstract} 
One of the most outstanding results of the \textit{Chandra} X--ray Observatory  was the discovery that AGN 
jets are bright X--ray emitters on very large scales, up to hundreds of kpc. 
Of these, the powerful and beamed jets of Flat Spectrum Radio Quasars are particularly interesting, 
as the X--ray emission cannot be explained by an extrapolation of the lower frequency synchrotron spectrum. 
Instead, the most common model invokes inverse Compton scattering of photons of the Cosmic Microwave Background 
(EC/CMB) as the mechanism responsible for the high energy emission. 
The EC/CMB model has recently come under criticism, particularly because it should predict a significant steady 
flux in the MeV--GeV band which has not been detected by the {\it Fermi}/LAT telescope for two of the best 
studied jets (PKS 0637--752 and 3C273). 
In this work we revisit some aspects of the EC/CMB model and show that electron cooling plays an 
important part in shaping the spectrum. 
This can solve the overproduction of $\gamma$--rays by suppressing the high energy end of the emitting 
particle population. 
Furthermore, we show that cooling in the EC/CMB model predicts a new class of extended jets that are bright 
in X--rays but silent in the radio and optical bands. 
These jets are more likely to lie at intermediate redshifts, and would have been missed in all previous 
X--ray surveys due to selection effects.
\end{abstract}


\section{Introduction}

One of the most remarkable achievements of the {\it Chandra} X--ray Observatory is the serendipitous 
discovery of bright, extended X--ray emission from large scale AGN jets, thanks to its exceptional 
angular resolution. 
Indeed, the first such jet was observed during the very first pointing of the telescope 
(Chartas et al. 2000, Schwartz et al. 2000).  
To date, dozens of these sources have been detected in a variety of AGN classes such as FRI and FRII radio 
galaxies, BL Lacs, as well as steep and flat spectrum radio quasars (see Harris \& Krawczynski 2006 for a review). 
A comprehensive catalogue can be found at {\tt https://hea-www.harvard.edu/XJET/}. 

Data gathered with {\it Chandra} over the years (Sambruna et al. 2004; Marshal et al. 2005; Hogan et al. 2011; 
Worrall et al. 2001; Marshall et al. 2011) has shown that the characteristics of the observed X--ray 
emission is strongly correlated with the jet's radio morphology. 
In low--power and/or misaligned sources such as radio galaxies (both of the FRI and FRII types), as well as 
some BL Lacs, the observed X--ray flux is usually consistent with an extrapolation of the non--thermal 
synchrotron emission responsible for the radio and optical data. 
On the other hand, the X--ray flux observed in the beamed and powerful extended jets of 
Flat Spectrum Radio Quasars (FSRQs) is several orders of magnitude above such an extrapolation, 
and therefore in these jets a second radiative component is required. 
The origin of this second radiative component is still debated.

Both SSC and bremsstrahlung are easily ruled out due to unphysical requirements (Chartas et al. 2000). 
Therefore, in analogy with the standard leptonic model for the core region of radio--loud AGN, 
Tavecchio et al. (2000) and Celotti et al. (2001) proposed that the high energy emission is due to 
external Inverse Compton (EC) scattering of soft photons by the same population of electrons responsible 
for the low--frequency synchrotron emission. 
For extended jets hundreds of kiloparsecs away from the host galaxy, the only viable target photon field 
is the Cosmic Microwave Background (CMB). 
The EC/CMB model has been used to successfully fit the SEDs of several FSRQ jets (Tavecchio et al. 2000, 
Celotti et al. 2000, Sambruna et al. 2004, Jorstad and Marscher 2004, Cheung 2004, Cheung et al. 2006, 
Sambruna et al. 2006, Simionescu et al. 2016), and is considered the standard radiative model for large 
scale jets generated in powerful AGN. 
Despite this, it has come under some criticism (Uchiyama et al. 2005, Jester et al. 2006, 
Uchiyama et al. 2006, Meyer and Georganopoulos 2014, Meyer et al. 2015), and in some of these 
works alternative radiative models have been proposed.

A key requirement of the EC/CMB model is that the jet needs to remain highly relativistic ($\Gamma \approx 10$) 
on very large scales; in this case, the energy density of the CMB is boosted by a factor $\Gamma^{2}$, making 
EC/CMB an efficient radiative mechanism capable of reproducing the bright fluxes observed in the {\it Chandra} band. 
The requirement of a highly relativistic jet naturally implies strong relativistic beaming ($\delta \approx 10$). 
Therefore, in this paper we only focus on the jets of FSRQs, and ignore radio galaxies and SSRQs, for 
which strong beaming is not expected.

The goal of this paper is to 
re-examine some aspects of the EC/CMB model, in order to reconcile it with recent 
data and to provide 
predictions testable with current and upcoming instruments. 
The paper is structured as follows: 
in \S 2 we review the state of the EC/CMB model, and discuss possible alternatives to it; 
in \S 3 we illustrate our updated EC/CMB model; 
in \S 4 we apply our new model to two notable sources; 
in \S 5 we discuss the implications of our new model, and 
in \S 6 we draw our conclusions.

Throughout the paper, the following cosmological  parameters are assumed:
$H_0=70$ km s$^{-1}$ Mpc$^{-1}$, $\Omega_{\rm M}=0.3$, $\Omega_{\Lambda}=0.7$.

\section{State of the EC/CMB model}

Despite its early success and general acceptance, in recent years the EC/CMB model has come under 
criticism from several authors. 
The two main issues raised are the diffuse emission problem (Tavecchio and Ghisellini 2003) and 
the missing $\gamma$--ray problem (Meyer and Georganopolous 2014, Meyer et al. 2015); further arguments 
against the EC/CMB model come from in--depth study of individual sources that show complex behavior 
(in particular 3C273, Jester et al. 2006; Uchiyama et al. 2006; and PKS1136--135, Sambruna et al. 2004; 
Sambruna et al. 2006; Cara et al. 2013), for which X--ray spectral data and optical polarimetry shows 
that a significant part of the high energy flux is produced by electron synchrotron emission. 
Notably, such behavior is not seen in other well studied sources like PKS 0637--752 and PKS 1150+497 
(Chartas et al. 2000; Mueller and Schwartz 2009; Sambruna et al. 2006).

A point in favour of the EC/CMB model comes from the recent serendipitous discovery of a bright 
X--ray jet in the intermediate redshift ($z=2.5$) quasar B3 0727+490. 
Simionescu et al. (2015) reported the detection of the jet with the {\it Chandra} telescope, despite 
the lack of an extended radio counterpart. 
Such behavior is fairly intuitive in the context of an External Compton model (Schwartz 2002), 
as the CMB energy density strongly increases with redshift, and therefore we expect unusually 
bright X--ray jets at higher redshifts ($z>2$) which are not observed in the current sample of closer sources ($z<1$).

\subsection{The diffuse emission problem}

The diffuse emission problem results from considering particle cooling in the emitting knots 
(Tavecchio and Ghisellini 2003). 
If the X--ray flux is produced by EC/CMB, the electrons up scattering the CMB need to have low 
energies ($\gamma \approx 10^{2}$), much lower than those producing the observed radio 
($\gamma \approx 10^{3}$) or IR-optical emission ($\gamma \approx 10^{5}-10^{6}$). 
Because the radiative cooling time scale depends on electron energy ($t_{\rm rad} \propto 1/\gamma$), 
we would expect to observe different morphologies in different bands. The IR and optical emission, 
produced by high energy electrons, would come from very compact zones, while radio and especially 
X--ray emission should come from more diffuse regions. 
We would also expect to observe relic knots that are leftovers of aged particle populations, and 
therefore bright in X--rays but faint in radio and optical bands. 
Such a behavior hasn't been observed; 
available radio, optical and X--ray data generally shows that the emission regions looks remarkably 
similar at very different wavelengths (Sambruna et al. 2004; Sambruna et al. 2006; Marshall et al. 2005). However, this could be caused by selection bias in the current sample of extended jets with available X-ray data (see Sec. 5 for further discussion).
A small amount of diffuse X--ray emission unrelated to the radio knots is sometimes observed 
(Sambruna et al. 2004; Sambruna et al. 2006; Schwartz et al. 2006; Perlman et al. 2011), but such detections are somewhat tentative except for two jets (Jorstad and Marscher 2004; Simionescu et al. 2016); typically the emission regions in the radio and X--rays appear very similar.

\begin{figure}
\vskip -0.7 cm
\hspace*{-0.8 truecm}
\includegraphics[scale=0.42]{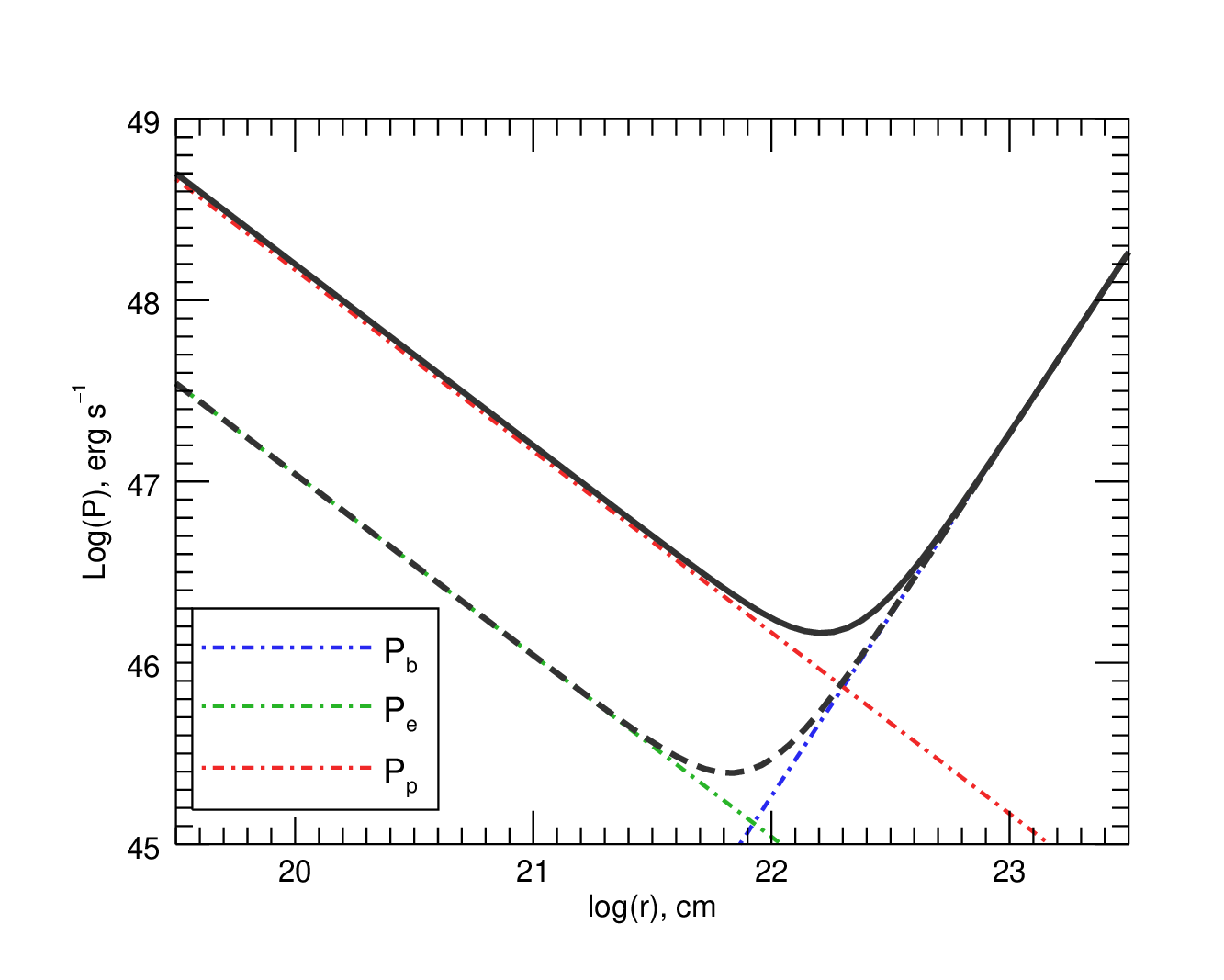}
\vspace{-0.6 cm}
\caption{
Knot power as a function of radius, assuming constant Lorentz 
factor ($\Gamma = 10$) and magnetic field ($B=10$ $\mu$G). 
The blue, green and red dash-dotted lines correspond to the power carried by electrons, protons 
and magnetic field, respectively. 
The black lines represent the total power in a single knot including the proton rest-mass energy density 
(thick black line) or excluding it (dashed black line). 
Smaller radii correspond to extremely high power.
}
\label{power}
\end{figure}

A possible solution is the existence of unresolved clumps in the jets (Tavecchio \& Ghisellini 2003). 
If the emission region is a few orders of magnitude smaller than the $\approx 10^{22}$ cm size 
typically assumed, and rapidly expanding, then adiabatic cooling can dominate over radiative losses. 
Adiabatic cooling is independent of particle energy, and therefore electrons with different Lorentz 
factors can cool at nearly identical rates. 
As a result, jet knots would have the same size in every observed band. 
However, we find this explanation to be unsatisfactory due to energy budget requirements. 
Jet power is typically computed for each knot as $P_k=R^{2}\Gamma \beta c (U^\prime_b+U^\prime_e+U^\prime_p)$, 
where $R$ is the radius of each emitting region, and $U^\prime_b$, $U^\prime_e$, $U^\prime_p$ 
are the energy densities of the magnetic field, electrons and protons, respectively (Celotti and Fabian 1993). 
One cold proton per electron is assumed. 
For a given magnetic field and bulk Lorentz factor, we need the same total number of
particles in order to produce the observed synchrotron and EC luminosities.
Therefore, by changing the size of the emitting region, we change the particle density
(but not the magnetic energy density, assumed to be constant).
If the emitting region is assumed spherical, we have that $U_e\propto  R^{-3}$, and thus
$P_e\propto R^2U_e\propto R^{-1}$.
The magnetic power is instead always $P_B\propto R^2$.
Fig. \ref{power} illustrates this point.
Small emitting regions (for example, $R \approx 10^{20}$ cm rather than $R \approx 10^{22}$ cm) correspond to higher kinetic powers. The same argument also holds if each knot contains several clumps instead of one:  the electron energy density necessarily increases because the same number of electrons is concentrated in a smaller volume. However, regardless of the number of clumps the number of particles is constant, and so their contribution to the jet's energy remains unchanged. Therefore the \textit{average} power of the jet (energy per unit of time) remains constant.

\subsection{The $\gamma$--ray problem}

The $\gamma$--ray issue originates from recent  {\it Fermi}/LAT data. 
The low magnetic fields invoked by the EC/CMB model require high energy electrons 
($\gamma\sim 10^5-10^6$)
to reproduce the observed optical flux; these electrons should scatter CMB photons to GeV energies
($\nu_{IC} \sim \gamma^2\nu_{\rm CMB}\delta \sim 10^{22} -10^{24}$ Hz).
The EC/CMB model therefore predicts that extended jets should be fairly powerful GeV emitters. 
However the extended jet emission is still 1--2 orders of magnitude more faint than the bright quasar 
core, and the {\it Fermi}/LAT telescope does not have the angular resolution necessary to resolve the 
two components. 
Therefore, timing--based techniques are required: the bulk of the emission originating from blazars 
comes from small emission regions, and is therefore highly variable; on the other hand, the knots 
observed likely originate in large emitting regions, and therefore they are not expected to show 
significant variability on time scales of a few years. 
In their recent work, Meyer and Georganopolous (2014) and Meyer et al. (2015) analyzed the 6--year 
light curves from the 3FGL catalogue of two FSRQs: PKS0637--752 and 3C273. 
For both objects they claim that {\it Fermi} did not detect a steady emission component, 
presumably associated with the extended jet, and that the available upper limits to this 
emission are lower than the predictions of EC/CMB, thus ruling it out. 
This is a rather strong point against the standard EC/CMB model, particularly in the case of 
PKS 0637--752, which historically has been considered the benchmark extended FSRQ jet.

In light of these findings, the interest in alternative emission models has increased. 
The two favoured candidate models invoke a more complex synchrotron model for the high energy component, 
either due to radiative cooling (Dermer and Atoyan, 2002) from a separate population of electrons 
(Uchiyama et al. 2005, Uchiyama et al. 2006, Jester et al. 2006), or from very energetic protons (Aharonian, 2002). 
These models can satisfactorily reproduce the observed SEDs, but suffer from two drawbacks: 
the presence of a second population of emitting particles is a rather ad--hoc assumption, 
and they are incapable of producing predictions that can be tested in the short term with 
current or upcoming instruments (Meyer et al. 2015). 
Nevertheless, they remain a viable alternative.

\section{EC/CMB revisited}

Historically, the SED of extended jets has been computed by assuming an ad--hoc shape for the 
electron distribution (either a single or a broken power law), in order to 
reproduce the observed fluxes and spectral indices with reasonable physical parameters 
(e.g. Sambruna et al. 2006). 
In this work instead we compute the evolution of an injected electron distribution by 
solving the continuity equation, accounting for both radiative and adiabatic cooling, in order 
to study the evolution of extended jets as a function of both time and redshift. 
This approach leads to more robust and consistent estimates of key quantities, such as the synchrotron 
peak luminosity and frequency. 
For simplicity we assume that the entirety of the observed emission 
comes from a single spherical ``blob" of radius $r$, moving with a bulk Lorentz factor $\Gamma$ and expanding adiabatically with a speed $\beta_{\rm exp}c$.

We define $N(\gamma,r)$ as the number of electrons with a given Lorentz factor $\gamma$, contained in a spherical blob of radius $r$ and with an initial radius $r_0$. 
The blob is assumed to be moving with a bulk Lorentz factor $\Gamma$. 
The evolution of the particle distribution is given by the continuity 
equation (Sikora et al. 2001):
\begin{equation}
\frac{\partial N(\gamma,r)}{\partial r} = 
Q\left(\gamma\right) - \frac{\partial}{\partial \gamma} \left(N \frac{d \gamma}{d r} \right),
\label{ngamma}
\end{equation}
where $Q\left(\gamma\right)$ is the particle injection term and $d\gamma / dr$ the energy loss term.
In this form, time has been replaced by the distance travelled: $r=\beta c \Gamma t^\prime$, where 
$t^\prime$ is the time measured in the comoving frame (Tavecchio and Ghisellini 2003).

The electrons lose energy through both adiabatic and radiative losses. EC/CMB is the dominant mechanism for radiative losses, and synchrotron cooling is negligible. This can be shown by comparing the magnetic and CM energy densities in the comoving frame of the jet:
\begin{equation}
U_{\mathrm{CMB}}(1+z)^{4}\Gamma^{2} = \frac{B^{2}}{8\pi},
\end{equation}
which can be inverted to write the CMB energy density in terms of an equivalent magnetic field:
\begin{equation}
B_{\mathrm{CMB}} = \left(8\pi U_{\mathrm{CMB}}\right)^{1/2} = 3.26 \cdot 10^{-6}(1+z)^{2} \Gamma\ \mathrm{G}.
\end{equation}

For typical parameters ($\Gamma = 10$, $z = 0.5$), $B_{\rm CMB}$ is roughly 1 order of magnitude higher than the magnetic field inferred from fitting the synchrotron part of the SED. Therefore, Compton losses dominate over sychrotron ones. The timescales for radiative and adiabatic losses are given by:
\begin{align}
& t_{\mathrm{rad}} = 
\frac{3m_{e}c^{2}}{4\sigma_{\mathrm{T}}cU_{\mathrm{CMB}}\Gamma^{2}\gamma\left(1+z\right)^{4}}  = \\ 
& 1.4 \times 10^{3} \left(\frac{10}{\Gamma}\right)^{2}\left(\frac{10^{6}}{\gamma}\right)\left(\frac{2}{1+z}\right)^{4}\ 
\mathrm{yr}
\label{rad}
\end{align}

\begin{align}
t_{\mathrm{ad}} = \frac{r}{\beta_{\mathrm{exp}}c} = 
3.2\times 10^{4} \left(\frac{0.1}{\beta_{\mathrm{exp}}}\right)\left(\frac{r}{\mathrm{kpc}}\right)\ \mathrm{yr}.
\label{ad}
\end{align}

The matching point between the two regimes, where the electron population is expected to change 
slope, is found by comparing the two time scales:
\begin{align}
& \gamma_{\mathrm{break}} = 
\frac{3}{4}\frac{\beta_{\mathrm{exp}}m_{e}c^{2}}{r\sigma_{\mathrm{T}}\Gamma^{2}U_{\mathrm{CMB}}(1+z)^{4}} = \\ 
& 4.5\times 10^{4} \left(\frac{1\mathrm{kpc}}{r}\right)
\left(\frac{\beta_{\mathrm{exp}}}{0.1}\right)\left(\frac{10}{\Gamma}\right)^{2}\left(\frac{2}{1+z}\right)^{4} 
\end{align}
%
The break Lorentz factor depends strongly on the source's bulk Lorentz factor, assuming a 
typical expansion speed and size. 
This can in principle help in breaking the typical degeneracy of the beaming factor between line of sight and bulk jet speed.

\begin{figure}
\vskip -0.5 cm
\hspace*{-0.8 truecm}
\includegraphics[scale=0.42]{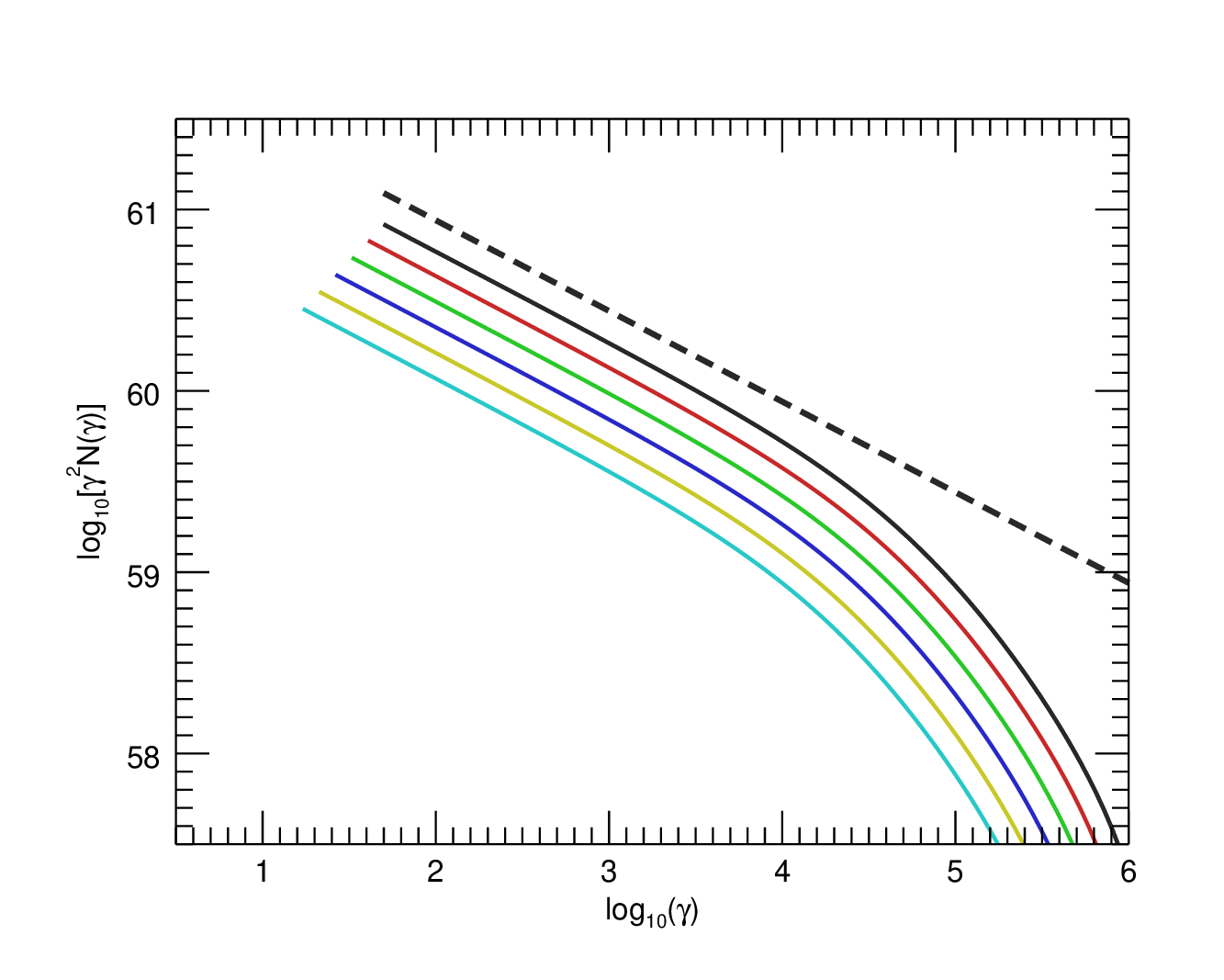}
\caption{
Particle distribution as a function of time,
according to the continuity equation (Eq. \ref{ngamma}), for typical jet parameters 
($B=10$ $\mu$G, $r_{0}=10^{22}$ cm,  $z=0.5$, $Q(\gamma) \propto \gamma^{-2.5}$, $\gamma_{\rm min}=50$). With such a choice of $\gamma_{\rm min}$, the optical emission is due exclusively to synchrotron.
The shown $N(\gamma)$ distributions are multiplied by $\gamma^{2}$ for illustrative purposes.
The dashed line shows the injected electron distribution; the colored lines show the evolved population. 
Each time--step corresponds roughly to one adiabatic time scale.}
\label{pop1}
\end{figure}
\begin{figure}
\hspace*{-0.8 truecm}
\includegraphics[scale=0.42]{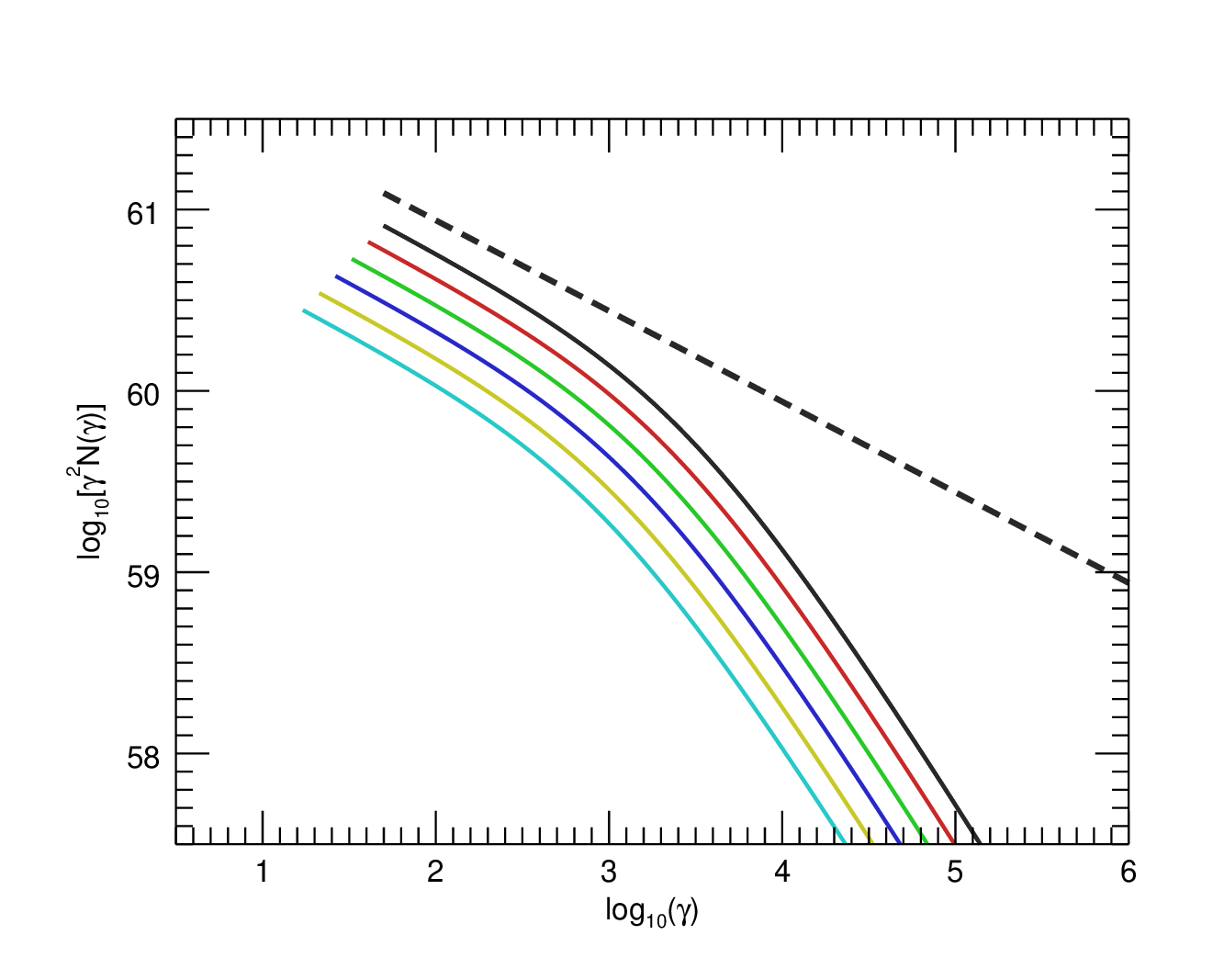}
\caption{
Evolution of the same particle distribution as in Fig. \ref{pop1}, but moved to redshift $z=2.5$. 
In this case, the effect of EC/CMB cooling of intermediate and high energy electrons is much more intense.
}
\label{pop2}
\end{figure}
\begin{figure}
\vskip -0.5 cm
\hspace*{-0.8 truecm}
\includegraphics[scale=0.42]{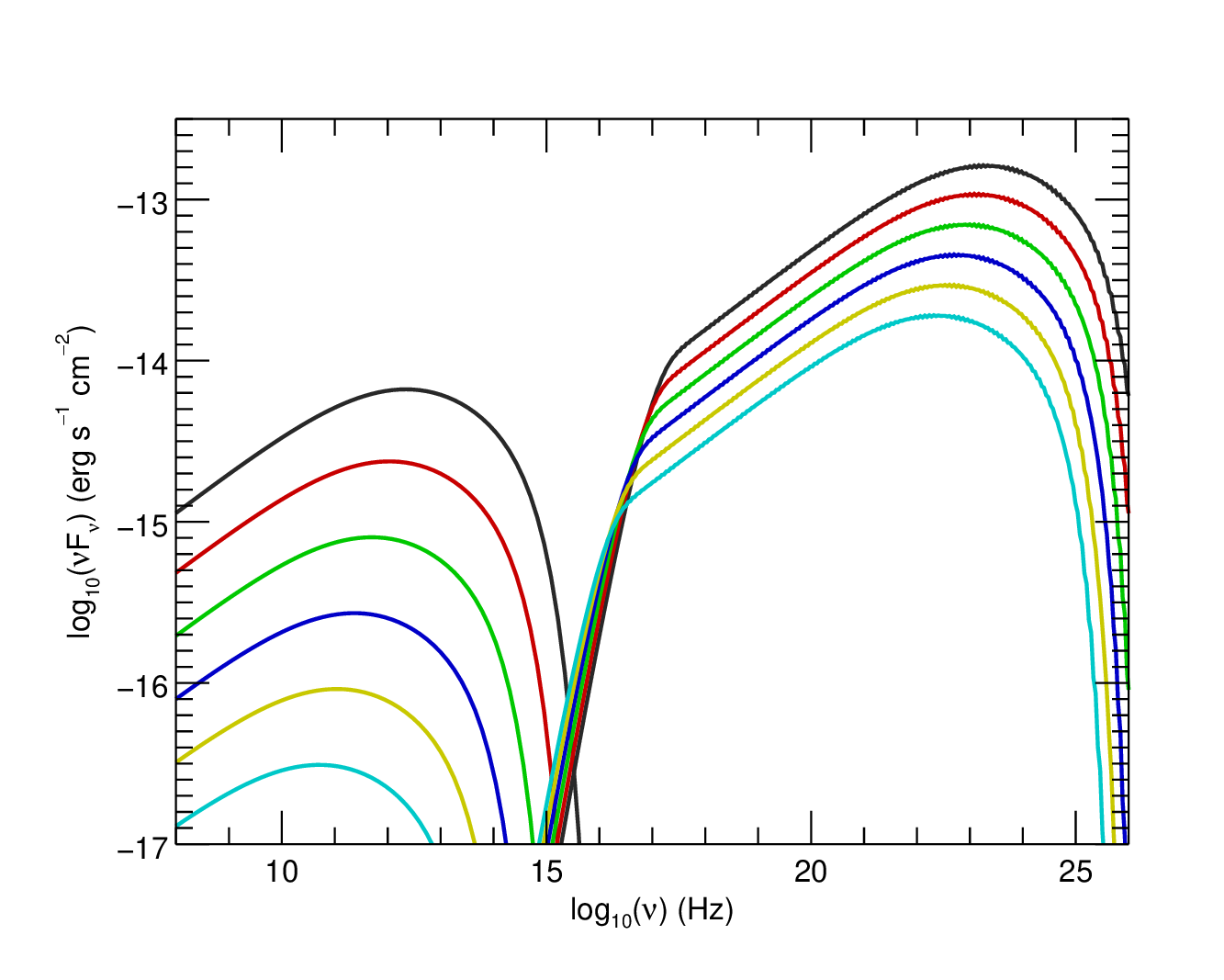}
\caption{
SED of the synchrotron+EC/CMB emission from the same parameters ($B_{0}=10$ $\mu$G, $r_{0}=10^{22}$ cm,  $z=0.5$) and particle distribution of Fig. \ref{pop1}, computed with the additional assumption that as the blob travels downstream, the magnetic field surrounding it decreases as $B(R)\propto R^{-1}$. 
This is the reason why the synchrotron flux decreases more rapidly than the inverse Compton one.
The optical synchrotron emission is the first to be suppressed by cooling, followed by IR, radio and X--rays in this order.}
\label{sed1}
\end{figure}
\begin{figure}
\vskip -0.4 cm
\hspace*{-0.8 truecm}
\includegraphics[scale=0.42]{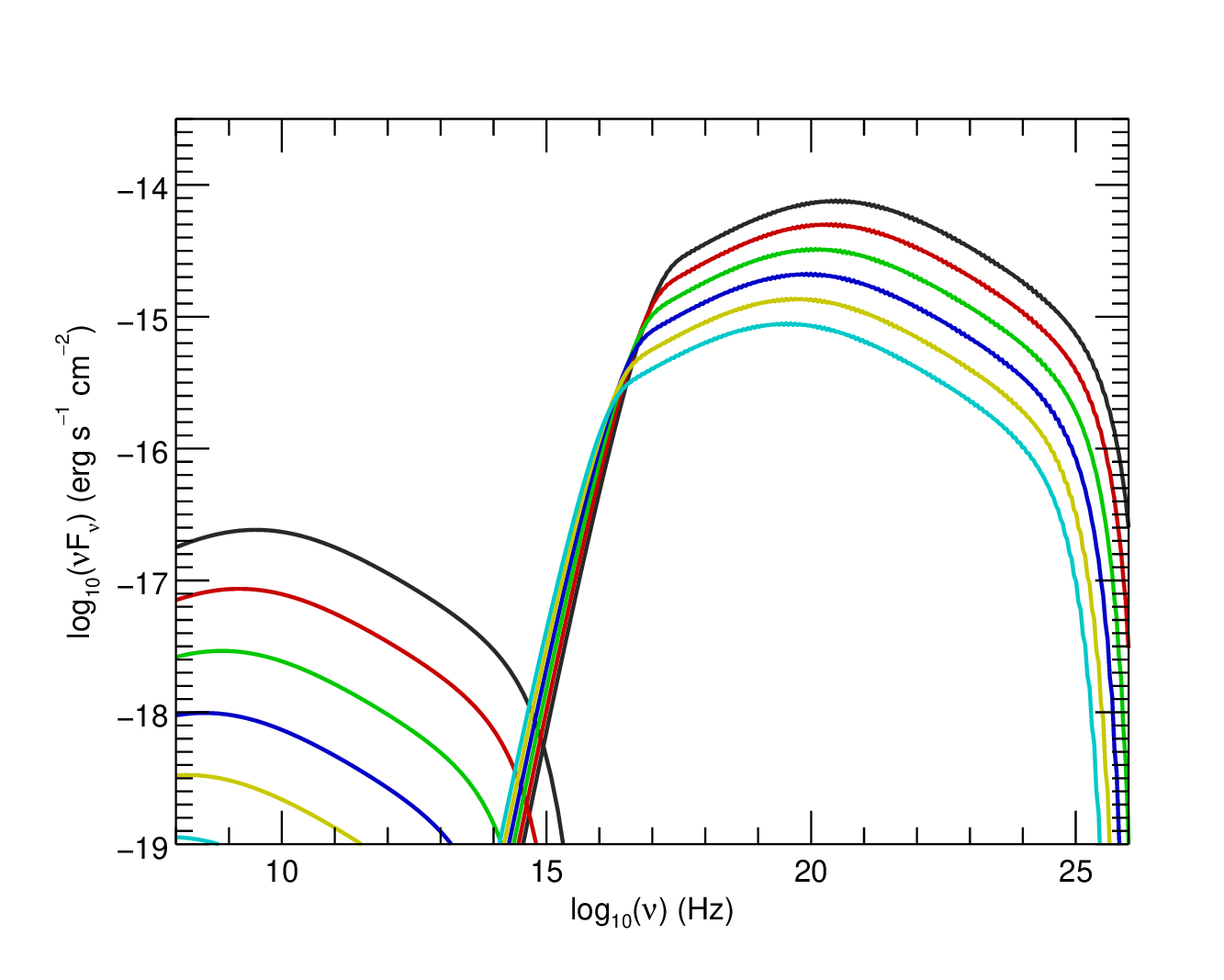}
\caption{
SED corresponding to the particle distribution of Fig. \ref{pop2}. 
The synchrotron emission is suppressed at all but the lowest frequencies, while EC/CMB emission at X--ray energies remains very bright.
}
\label{sed2}
\end{figure}

The solution of the continuity equation is, in the general case, time--dependent. 
However, eventually the particle distribution will reach a stationary state in which cooling and injection are balanced. 
In this state $N(\gamma)$ becomes time--independent and is given by:
\begin{equation}
0 = Q(\gamma) - \frac{\partial}{\partial \gamma}\left(N\frac{d\gamma}{dr}\right),
\label{steady}
\end{equation} 
yielding:
\begin{equation}
N(\gamma) \propto \int_{\gamma}\frac{Q\left(\gamma^{\prime}\right) d\gamma^{\prime}}{\dot{\gamma}}.
\label{steady2}
\end{equation}
The injection term is always assumed to be a power-law:
\begin{equation}
Q(\gamma) = \frac{K}{t_{\rm inj}} \gamma^{-s},
\end{equation}
where the injection time $t_{\rm inj}$ is always taken to be $r_{0}/c$ and the constant $K$ is the electron numerical density after $t_{\rm inj}$.

Once injection shuts down the solution is time--dependent; the continuity equation becomes (Tavecchio and Ghisellini, 2003):
\begin{equation}
\frac{\partial N}{\partial r} = \frac{\partial}{\partial \gamma}\left(N\frac{d\gamma}{dr}\right)
\end{equation} 
The cooling term as a function of $r$ has the form:
\begin{equation}
\frac{d\gamma}{dr} = -\frac{1}{\beta c \Gamma}\frac{4\sigma_{\mathrm{T}}cU_{\mathrm{CMB}}}{3m_{e}c^{2}}\gamma^{2} - 
\frac{2\gamma}{3r},
\end{equation}
where the first term is due to radiative cooling and the second to adiabatic losses. 
The solution for this equation is:
\begin{equation}
\gamma(r) = \frac{1}{3\left(\alpha r + (r/r_{0})^{-2/3}[
1/(3\gamma_{0})-\alpha r_{0}]\right)},
\end{equation}
where $r_{0}$ is the initial knot radius and 
$\alpha = 4\sigma_{\mathrm{T}}U_{\mathrm{CMB}}/\left(3\beta_{\mathrm{exp}}\Gamma m_{e} c^{2}\right)$. 
This means that all particles with initial energy between $\gamma_{0}$ and $\gamma_{0}+d\gamma$, 
after a time $t$ (or a distance $r$), will have energy between $\gamma$ and $\gamma+d\gamma$. 
Differentiating the solution we find:
\begin{equation}
\frac{d\gamma}{d\gamma_{0}} = \frac{1}{(3\gamma_{0})^{2}}\left(\frac{r}{r_{0}}\right)^{2/3}
\left[\alpha r +\left(\frac{r}{r_{0}}\right)^{2/3} \left(\frac{1}{3\gamma_{0}} - 
\alpha r_{0} \right)\right]^{-2}.
\end{equation}
The particle distribution at a given distance $r$ (corresponding to a given time $t$) is obtained 
by calculating the initial $\gamma_{0}$, cooled to $\gamma$, and taking into account the 
corresponding differentials:
\begin{equation}
N(\gamma,r) = \frac{N_{0}(\gamma_{0},r_{0})}{d\gamma/d\gamma_{0}}.
\end{equation}
The exact solution has to be obtained by numerical integration; solutions at different times and with representative values of the jet parameters are shown in Fig. \ref{pop1}. 

The effect of aging on the corresponding SED is shown in Fig. \ref{sed1}, which was computed 
making the additional assumption that as the blob moves downstream, the magnetic field 
surrounding it decreases as $B(R) \propto R^{-1}$, where $R$ is the distance along the jet from the AGN core. 
Because of radiative cooling, the synchrotron peak moves to lower frequencies over time, 
the high energy electrons disappear quickly, and the optical and $\gamma$--ray fluxes are 
the first to be suppressed, followed by mid--IR, radio and finally X--ray in this order. 
Therefore, CMB cooling can be a way of curing the $\gamma$--ray problem.

Radiative cooling becomes more important at higher redshifts, due to the increased CMB energy density. 
In Fig. \ref{pop2} and Fig. \ref{sed2} we show the evolution and SED of the same injected electron 
population as in Fig. \ref{pop1} and Fig. \ref{sed1}, but moved to redshift $z=2.5$. 
At this higher redshift the cooling break is moved to lower electron energies; therefore 
the synchrotron emission is greatly quenched, and the synchrotron peak moves to lower frequencies. 
Vice versa, the EC/CMB component remains extremely bright, both because the intrinsic luminosity is 
necessarily higher $\left[L_{\rm EC/CMB} \propto U_{\rm CMB} \propto (1+z)^4\right]$, and because we naturally 
expect a higher number of low energy electrons to be present. 
Also, the spectra of jet knots tends to be fairly hard, and therefore K--correction effects will also 
increase the observed X--ray flux in the 0.3-10 keV band. Additionally, the peak of the EC/CMB components moves from the GeV to the hard--X band.

This shows that the properties of extended jets emitting through EC/CMB are expected to change 
dramatically with redshift. 
Nearby sources ($z<1$) that are bright in radio should display significant X--ray emission, 
and sometimes optical emission as well. 
This is consistent with surveys conducted with {\it Chandra} (Marshall et al. 2005, Hogan et al. 2011) and the addition of {\it HST} (Sambruna et al. 2004), all of which have been based on radio criteria. 
Far away sources ($z>1$), for which available data is much more limited, should be roughly as bright 
as nearby ones in X--rays, at the expense of limited radio and especially optical emission.

\begin{figure}
\hspace*{-0.8 truecm}
\includegraphics[scale=0.42]{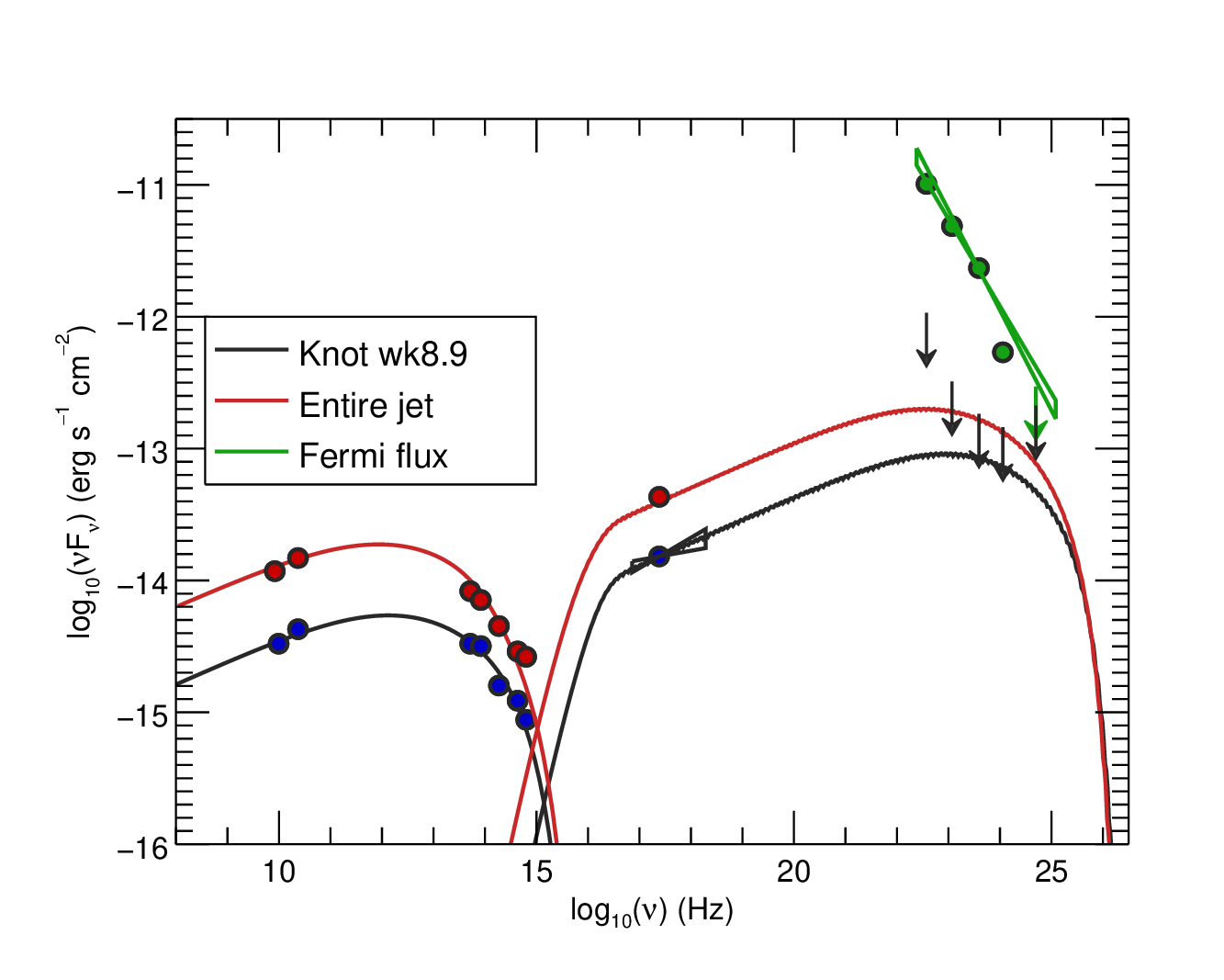}
\caption{
SED of the jet of the benchmark quasar PKS0637--752. 
The blue dots report the observed fluxes of the brightest jet knot (wk 8.9), the red dots 
the integrated flux of the entire jet. 
The black and red lines have been obtained with the updated EC/CMB model, which satisfactorily reproduces the radio, IR, optical and X--ray data while remaining below the available 
$\gamma$--ray upper limits. The black arrows are the {\it Fermi}/LAT upper 
limits for the steady extended component (from Meyer et al. 2015); the flux detected in each band by {\it Fermi}/LAT, as well as the power law fit reported in the 3FGL (Fermi-LAT Collaboration, 2015) is shown in green.
}
\label{0637}
\end{figure}

\section{The case of PKS 0637--752 and B3 0727+490 }

In this section we apply the updated EC/CMB to the available data from two notable sources: 
the benchmark jet of PKS 0637--752, and the newly discovered radio silent jet of B3 0727+490. 

We do not consider here other well studied jets such as 3C273 and PKS 1136--135 due to their unusual and complex behavior. In particular, in both sources the slope of the X--ray spectrum changes along the length of the jet, while the radio slope remains constant. Jester et al. (2006) showed that in the case of 3C273 this is inconsistent with a one--zone model, and a significant amount of the emission could originate in a slow-moving layer surrounding a fast relativistic spine. The jet of PKS 1136--135 behaves similarly (Sambruna et al. 2006), and additionally the source is classified as a steep spectrum radio quasar, possibly implying that the jet is fairly misaligned with respect to the line of sight. In this case, in the context of a spine-layer model scenario suggested for 3C273 we do not expect the spine to dominate the observed emission. We also note briefly that in more misaligned FRII radio galaxies such as Pictor A a second radiative component isn't required, and the bulk of the emission from radio to X--rays is generated through the synchrotron process by a mildly relativistic plasma (Hardcastle et al. 2016).
 
The parameters for the fits are reported in Table 1. 
Due to the limited number of data points and high number of free parameters (9), we do not perform 
goodness--of--fit tests, but instead use a ``fit--by--eye" approach, as is typically done in many similar studies.

\subsection{PKS 0637--752}

The benchmark jet of PKS 0637--752 is one of the two for which the $\gamma$--ray flux predicted by EC/CMB was reported to be higher than that measured by {\it Fermi}/LAT (Meyer et al. 2015). It is also  the jet with the highest quality data available, as it has been observed in the radio, X--ray, mid--IR and optical bands.

The result of fitting all the jet knots with the updated EC/CMB model, with the parameters listed in Table 1, is shown in Fig. \ref{0637}. Such parameters correspond to equipartition between the magnetic field and relativistic electrons. All knots are reproduced satisfactorily, but only the brightest (knot wk8.9) and the integrated flux of the jet are shown in figure for clarity. 

Mid-IR data from the Spitzer telescope shows that the synchrotron component follows a broken power law (Uchiyama et al. 2005), and the peak is well constrained by {\it HST} and {\it Spitzer} data. Our EC/CMB model is capable of reproducing such behavior .

Using our updated model, we take the injection term to be a power-law (eq. \ref{steady2}), and then consider the effects of cooling on such a particle distribution.  Because the jet shows bright synchrotron emission, we assume for simplicty that the emission of knots is produced by a stationary state particle population. 

Due to the short radiative time scale (eq. \ref{rad}), if we were to consider the case of later times after switching off the injection term, then the number of high energy electrons would be too low to reproduce a significant optical and IR flux. Vice versa, if we were to consider earlier times then the shape of the electron distribution would resemble that of the injection, which would result in an optical/IR flux brighter than the one observed. The solution to the continuity equation is therefore given by eq. \ref{steady}; we then compute the corresponding SED from the evolved particle distribution, which includes the effect of mild ($\beta < 1$) adiabatic cooling. 

The model well represents the entire synchrotron spectrum, and therefore provides a solid estimate for the $\gamma$--ray flux. In turn, the model also suggests that cooling is an important factor for shaping the overall SED: in this jet peak synchrotron emission is slightly suppressed due to cooling, and so is the corresponding $\gamma$--ray flux, which is generated by the same high energy electrons. Therefore, the updated EC/CMB model predicts a $\gamma$--ray flux {\it below} the {\it Fermi}/LAT upper limits in all bands, albeit barely. 

In Meyer et al. (2015) the only band in which the data suggests a possible detection of the extended jet is that  between 3 and 10 GeV.
Quite interestingly in this energy range our model predicts the largest contribution of the large scale knots with respect to the core flux. This band is therefore the first in which we expect to detect the knot: this would be a marginal, but promising, evidence in favour of our model.

We therefore conclude that current data cannot rule out the external Compton mechanism as the predominant radiative process for the high energy emission of the knots in PKS 0637--752, as the $\gamma$--ray flux predicted by EC/CMB is still beyond the capabilities of the Fermi/LAT telescope. 
Furthermore, our results highlight the importance of computing in detail the shape of the synchrotron peak to correctly predict the corresponding $\gamma$--ray flux.

\begin{figure}
\hspace*{-0.8 truecm}
\includegraphics[scale=0.42]{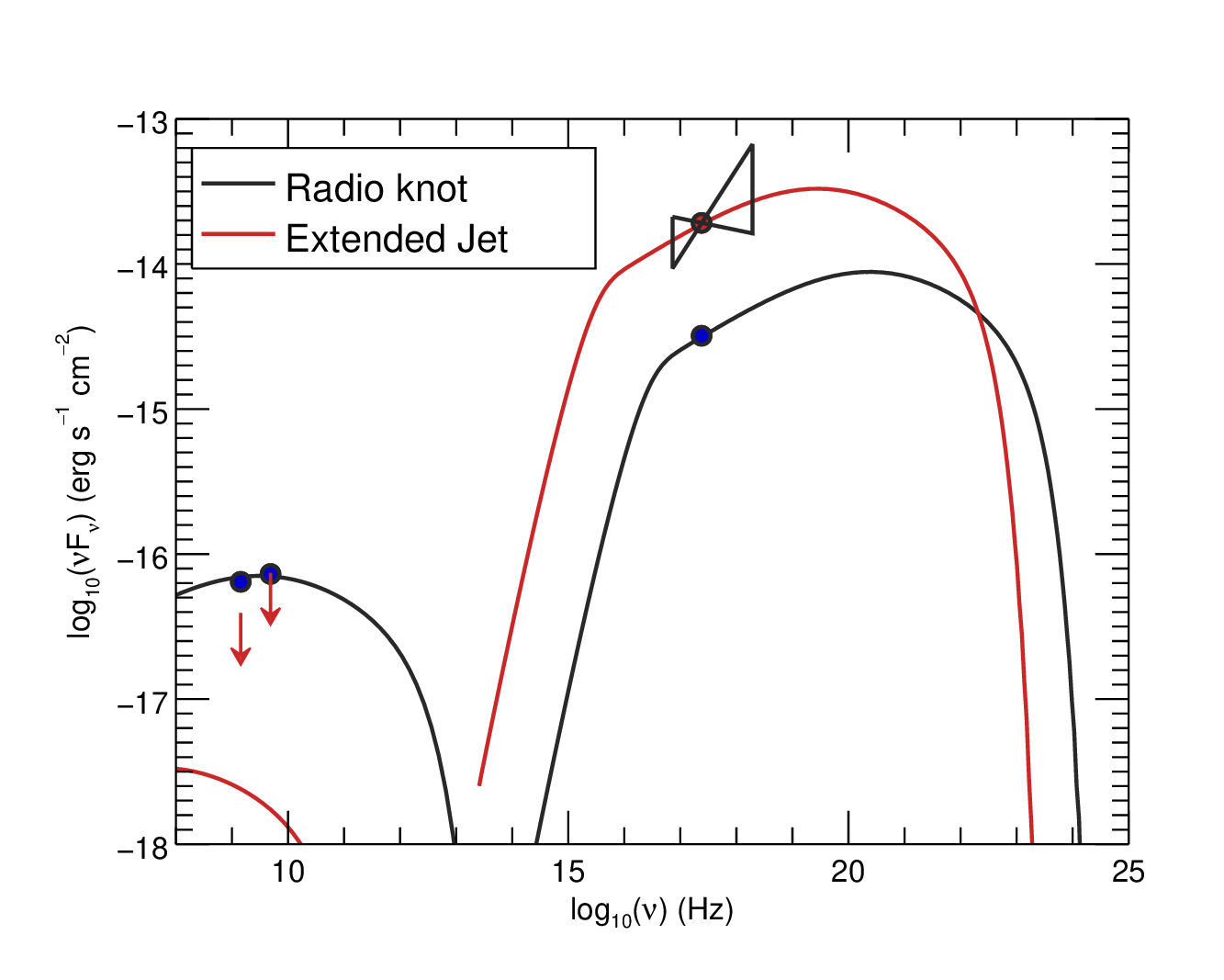}
\caption{
SED of the intermediate redshift ($z=2.5$), radio silent jet of B3 0727+409. 
The blue and red dots correspond to the observed fluxes of the radio-bright 
knot and X--ray bright extended jet, respectively. 
Upper limits to the radio emission of the extended jet are shown as red arrows. 
The black and red lines show the EC/CMB fit of the radio knot and extended, radio--silent component. }
\label{0727}
\end{figure}

\begin{table*}
\centering
\begin{tabular}{| l || c | c | c | c | c | c | c | c | c | c |}\hline
Name &$\gamma_{\rm min}$  &$\gamma_{\rm max}$  &$s$  &$B$  &$K$  &$r$ &$\delta$  &$\Gamma$  &$\theta$  &$\beta$\\
 &  &  &  &$\mu$G  &$10^{-6}$ cm$^{-3}$  &10$^{22}$ cm  &  &  & deg &  \\ 
\hline 
0637--752 wk5.7 &3  &$2.3\times10^{5}$ &2.65 &27 &25 &0.6 &11  &11 &5.2 &0  \\
0637--752 wk7.8 &20 &$1.8\times10^{6}$ &2.7  &15 &44 &1   &9.5 &10 &5.9 &0.22 \\
0637--752 wk8.9 &20 &$1.6\times10^{6}$ &2.7  &14 &44 &1   &9.7 &8 &5.7 &0.25 \\
0637--752 wk9.7 &12 &$1\times10^{5}$   &2.7  &17 &19 &1   &9   &8 &6.2 &0.1 \\
0727+490 knot   &25 &$1\times10^{5}$   &2.5  &14 &18 &1   &9   &12 &6.1 &0.3 \\
\hline  
\end{tabular}
\caption{
EC/CMB fit parameters for the knots in the jets discussed in this paper. We assume to inject a single power law of relativistic electrons, with a slope $s$; $Q(\gamma)\propto \gamma^{-s}$ between $\gamma_{\rm min}$ and $\gamma_{\rm max}$, and normalization $K$. The injection time $t_{\rm inj}$ is always assumed to be $r/c$;
$r$ is the assumed size of the emitting region. $B$ is the value of the magnetic field, assumed to be homogeneous and tangled
throughout the source. $\beta$ represents the velocity of the adiabatic expansion. $\theta$ is the angle of the jet with the line of sight.}
\label{table}
\end{table*}

\subsection{B3 0727+490}

This is perhaps the most interesting extended jet discovered so far, as its existence goes against the  common paradigm that these objects emit strongly in the  radio band.
The quasar core and a small knot near it are detected in both radio and X--rays, but the large scale X--ray jet, 
detected up to a projected distance of $80 \ \mathrm{kpc}$ from the core, lacks a radio counterpart. 
Both the inner radio--bright knot and the extended jet are resolved by {\it Chandra}. Simionescu et al. (2016), using a standard EC/CMB model, estimate the jet bulk Lorentz factor ($\Gamma \approx 10$), viewing angle ($\theta \approx 7^{\circ}$), and deprojected size for the entire jet ($L \approx 600 \mathrm{kpc}$). Such estimates are consistent with those obtained for lower redshift sources. Furthermore, the observed morphology supports the assumption that the bulk of the diffuse X--ray emission is originates in the entire jet volume rather than in the compact bright knots.

In the framework of our model, such a jet represents the prototypical ``cooled" EC/CMB source, and its discovery constitutes very strong 
evidence in favour of the EC/CMB scenario. 
The lack of extended radio emission implies that any similar source would not have been included in previous 
surveys, which generally start from samples of sources with bright radio jets (Sambruna et al. 2004, Marshall et al. 2005, Worrall et al. 2011). Additionally, the rather high redshift suggests that radiative cooling through the EC/CMB mechanism plays 
an important role. We fit the SED of both the radio-bright knot near the core, and the extended radio--silent jet, shown in Fig. \ref{0727}. 

The emission of radio--bright knot is well reproduced by a set of parameters typical of such extended jet knots and assuming equipartition between the emitting electrons and the magnetic field energy densities. In this sense, the radio--bright knot is very similar to its low redshift counterparts.

In order to estimate the emission of the extended radio--silent jet, we assume it originates from the evolution of several blobs distributed along the jet. We take each ``initial blob" to have the same parameters as the inner radio--bright knot, and then let them evolve for several adiabatic time scales. Because the adiabatic timescale is by definition longer than the injection time scale (eq. \ref{ad}), we assume that injection starts at the same time in all the blobs.

Ageing the knots has deeply different effects on the two radiative components. 
The synchrotron emission decreases, as the intermediate energy electrons responsible for 
it cool through adiabatic and radiative losses, while the magnetic field decreases as 
the emitting blobs travel along the jet length. On the other hand, during ageing of all the knots the X--ray Compton component becomes {\it brighter} because the many relic knots along the jet only contain low energy electrons after cooling. Note that we are comparing several aged knots (in the extended jet) to a single young knot near the core. In this case, the integrated emission from all the aged knots becomes faint (below detection threshold) in the radio band, but it is brighter than the inital emission of a single knot in the X-ray band.

\section{Discussion}

We have shown that evolution of the particle energy distribution of the emitting electrons can have a key role in shaping the resulting synchrotron and EC/CMB components. In fact, the effects of cooling in the particle population can account for both the diffuse emission and 
the missing $\gamma$--ray issues that have been raised against the EC/CMB model. 

As the SED of PKS 0637--752 shows, the flux of the external Compton component in the {\it Fermi}/LAT is tightly constrained by available optical and IR data on the synchrotron peak, as both are generated by the same electrons. If as data suggests the particle population shows signs of cooling, then the total number of high energy electrons responsible for optical/IR and $\gamma$--ray emission is lower. This effect is most noticable around the peak synchrotron and inverse Compton peaks, both of which are lower than in the case without cooling.

In light of this, we suggest that solid criteria to distinguish candidate $\gamma$--ray bright extended jets from fainter sources are 
a) the jet knots display bright optical and IR emission in several bands 
b) the overall spectrum of most knots is fairly hard in the optically thin radio and the X--ray bands ($\Gamma_{X} \approx 1.7$).
 
These criteria correspond to young sources in which radiative cooling is negligible (a) and for which the bulk of the X--ray emission can be produced by low energy electrons, which are expected to follow the same distribution as those emitting in the radio band (b). Out of every extended jet studied so far, the only source that matches these criteria is the jet of PKS 1150+497 (Sambruna et al. 2006). 
Most of the other observed sources only have one or two measurements in the IR--optical--UV bands, and therefore the peak synchrotron emission is not well constrained. As {\it Fermi}/LAT gathers more data, further observations  with {\it HST} and (in the future) with JWST will be crucial in identifying $\gamma$--ray bright candidates.

Multi--wavelength data on extended quasar knots are currently limited 
to relatively nearby jets ($z<1.5$) selected through radio criteria (Sambruna et al. 2004, Marshall et al. 2005, Hogan et al. 2011). 
In this kind of sources, cooling is only important for the high energy end of the particle 
population, while the radio--synchrotron emission has not been suppressed by energy losses yet. 
This is consistent with the 40\% optical detection rate of jets that are bright in both  radio and X--rays (Sambruna et al. 2004). 
In the context of EC/CMB models such radio--bright jets are therefore fairly young, 
and we would not expect relic knots (i.e. with X--ray, but no radio emission) to be commonly observed. 
Indeed, only one jet (0827+243, Jorstad and Marscher 2004) displays X--ray bright, radio--silent knots. Due to all these considerations, we propose that the diffuse emission problem can be entirely due to selection effects in the surveys conducted thus far.

The main result of this paper is the prediction of the existence of a new class of radio--silent extended quasar jets that have been missed in observations conducted thus far due to selection bias. The prototype of such a source is the jet of B3 0727+490, whose emission can be satisfactorily reproduced by computing electron cooling in an EC/CMB model. 
Due to the strength of radiative cooling, radio-silent jets are expected to be very common  and easily detectable with {\it Chandra}, particularly at intermediate redshifts. 

Such a prediction greatly differs from the expectations of alternative models that invoke synchrotron as the dominant mechanism for X--ray emission in extended quasar jets.  Because of radiative cooling, the extremely high energy electrons responsible for 
synchrotron X--rays should cool on timescales much shorter than those producing radio emission: therefore, far away ($z\geq 2$) jets are expected to either be completely undetectable at all frequencies (if they are old sources), or at most bright in radio but faint in X--rays (if the emitting particle population is younger). 
Such radically different behaviors offer a very strong diagnostic to distinguish between the radiative models proposed so far, which are all capable of reproducing the SEDs of nearby, radio--bright jets.

\section{Conclusion}

In this paper we have briefly reviewed the state of the standard model for the broad band 
emission of powerful, extended jets in beamed sources (Flat Spectrum Radio Quasars), which 
invokes inverse Compton scattering with CMB photons in order to reproduce the observed X--ray fluxes. 
The two main issues raised with the standard model are the lack of $\gamma$--ray detections by the {\it Fermi}/LAT  
telescope (which we call the missing $\gamma$--ray problem), and the apparent similar size of the 
emitting regions in the radio, X--ray and optical bands (the diffuse emission problem). 
We find that the proposed solution of Tavecchio and Ghisellini (2003) to the diffuse emission problem 
leads to unphysical powers for the jet knots, and is therefore an unlikely scenario.

In order to account for both issues we proposed an extension of the standard EC/CMB model, in which 
the particle population responsible for the observed emission is computed self--consistently through 
the continuity equation in order to correctly account for radiative and adiabatic cooling. 
We find that our approach can both solve the missing $\gamma$--ray issue for the benchmark extended 
jet in the FSRQ PKS 0637--752, and can reproduce the SED of the peculiar radio-silent 
jet of the intermediate redshift ($z=2.5$) FSRQ B3 0727+490. 
Furthermore, we propose that the diffuse emission problem is a product of selection bias 
favouring younger sources. 

Our updated model also produces a very strong testable prediction on a new class of sources, 
which have been missed in surveys conducted thus far. 
Due to the effects of radiative cooling we expect a large amount of extended radio--silent jets 
that display significant amounts of X--ray emission, particularly at intermediate and high 
redshifts thanks to the increased CMB energy density and K--correction effects. 
Popular alternative models to EC/CMB can very easily reproduce the observed SEDs 
of nearby ($z\leq 1$) sources, but they predict an opposite trend of very limited 
X--ray luminosity at higher redshift. 
Therefore, particle cooling leads to a very powerful diagnostic for broad band radiative models in these sources.

\section*{Acknowledgments}
We are grateful to the anonymous referee for a careful reading and useful comments. We thank E. Meyer for useful comments. This work has been partly founded by a PRIN-INAF 2014 grant.

\end{document}